\documentclass[10pt,a4paper]{article}

\usepackage{jheppub}

\usepackage{color}
\usepackage{colordvi}
\usepackage{changes}

\usepackage{epsfig,epsf}
\usepackage{amsmath}
\usepackage{amsthm}
\usepackage{amsfonts}
\usepackage{amssymb}
\usepackage{dsfont}
\usepackage{epstopdf}
\usepackage{multirow}
\usepackage{braket}
\usepackage{tabularx}

\usepackage{rotating}

\usepackage{marvosym}

\usepackage{todonotes}

\usepackage{subcaption}
\usepackage{array}   
\newcolumntype{C}{>{$}c<{$}}

\usepackage{slashed}

\usepackage[active]{srcltx}

\newcommand{\s}[1]{\slashed{#1}}

\newcommand{\w}[1]{\widetilde{#1}}
\newcommand{\f}[1]{\mathcal{#1}}

\newcommand{\bb}[1]{\mathbb{#1}}

\def\II{\hbox{{1}\kern-.25em\hbox{l}}}

\def\II{\hbox{{1}\kern-.25em\hbox{l}}}

\title{
Threshold resummation for double-deeply virtual Compton scattering}

\author[a]{J. Schoenleber}
\affiliation[a]{
   RIKEN BNL Research Center, Brookhaven National Laboratory, Upton, NY 11973, USA}

\emailAdd{jschoenle@bnl.gov}

\abstract{
The threshold region for double-deeply virtual Compton scattering (DDVCS) is discussed.
I derive a resummation formula for the (partonic) threshold logarithms in the flavor non-singlet case. 
The resummations can be done by using (re)factorization theorems for the coefficient functions near the partonic thresholds. As a byproduct, we obtain the leading term in the threshold limit of the two-loop coefficient function in double-deeply-virtual Compton scattering, which agrees with the recent result from explicit calculation, providing a highly non-trivial cross-check. 
}


\begin{document}
\maketitle

\section{Introduction and summary of main results}\label{sec:introduction}

An indispensable tool for perturbative QCD is the resummation of logarithms through solving evolution equations. This precedes an all-order factorization of the quantity in question into two (or more) factors, in the limit where the logarithm of the ratio of two (or more) variables becomes large. The all-order separation of scales then allows one to predict certain terms to all orders using only fixed order results. In the best case, one can, in the limits where ratios of scales are large, reduce everything to single-scale problems.
In this work we will peform such an analysis for the virtual Compton process
\begin{align}
\gamma^{(*)}(q) N(p) \rightarrow \gamma^{(*)}(q') N(p'),
\label{eq: VCS}
\end{align}
where at least one of the photons is off-shell and its virtuality is much larger the target mass $m^2 = p^2 = p'^2$, the QCD scale $\Lambda_{\rm QCD}$, and the squared momentum transfer $-t = -(p-p')^2 > 0$.
For simplicity, we will throughout this work consider the case where \textit{both} $q$ and $q'$ are space-like (including the possibility that $q'$ is light-like). This is different from the practically relevant processes of double-deeply virtual Compton scattering (DDVCS) \cite{Muller:1994ses}, where $q^2 < 0, \, q'^2 > 0$. However, all the results can be analytically continued to recover this case. Note that, in spite of $q'$ being space-like, we will still refer to the case where two photons are far off-shell as DDVCS. The choice of taking $q'$ space-like is convenient, because we want to understand the connection between the threshold resummation of deep-inelastic scattering (DIS), where $q^2 = q'^2 < 0$, and deeply-virtual Compton scattering (DVCS) \cite{Ji:1996ek, Ji:1996nm}, where $q^2 < 0, \, q'^2 = 0$.

The amplitude for the process in eq. \eqref{eq: VCS} factorizes in terms of  hard coefficient functions (CFs) and generalized parton distributions (GPDs). In this work, we focus on the flavor non-singlet contribution to the leading power in $\frac{m}{Q}, \frac{\Lambda_{\rm QCD}}{Q}, \frac{\sqrt{-t}}{Q}$, which are given by the vector and axial-vector amplitudes for transversely polarized virtual photons
\begin{align}
\f F_{\perp}^{\rm ns} = \int_{-1}^1 \frac{dx}{\rho} \, C(x/\rho, \xi/\rho, \mu,Q) F_q(x,\xi,t,\mu) \notag
\\
\w { \f F }_{\perp}^{\rm ns} = \int_{-1}^1 \frac{dx}{\rho} \, \w C(x/\rho, \xi/\rho, \mu,Q) \w F_q(x,\xi,t,\mu)
\label{eq: main factorization DDVCS}
\end{align}
where $\rho$ is the generalization of Bjorken-$x$, $\rho |_{q = q'} = x_B \equiv \frac{Q^2}{2p \cdot q}$ (see eq. \eqref{eq: kin defs} for the precise definition), $\xi$ is the skewness, $t$ is the squared momentum transfer, and $Q^2 = - \frac{1}{4} (q+q')^2$ is the hard scale\footnote{Note that this differs by a factor of $2$ from the notation usually used in DVCS, i.e. $Q_{\rm DVCS}^2 = -q^2 = 2 Q^2 |_{q'^2 = 0}$.}. The factorization in eq. \eqref{eq: main factorization DDVCS} separates the hard scale $Q^2$ from the small scales $\sqrt{-t}, m, \Lambda_{\rm QCD}$. $F_q, \, \w F_q$ are the unpolarized and polarized quark GPDs respectively and $C,  \w C$ are the corresponding flavor non-singlet CFs.

In the following, we give results for $\f F_{\perp}^{\rm ns}$ (or $C$ equivalently) but we will find that the same results hold for $\w{\f F}_{\perp}^{\rm ns}$ (or $\w C$ equivalently). The overall normalization is such that
\begin{align}
C &= \frac{\rho}{\rho - x - i0} - \frac{\rho}{\rho + x - i0} + O(\alpha_s),
\\
\w C &= \frac{\rho}{\rho - x - i0} + \frac{\rho}{\rho + x - i0} + O(\alpha_s).
\end{align}
While $m \sim \Lambda_{\rm QCD}$, the presence of the additional external scale $\sqrt{-t}$  suggests the interesting kinematic region where $Q \gg \sqrt{-t} \gg \Lambda_{\rm QCD}$. Indeed, in this limit the quark GPD admits a factorization separating the scales $\sqrt{-t}$ and $\Lambda_{\rm QCD}$ \cite{Diehl:1999ek, Vogt:2001if, Hoodbhoy:2003uu}. However, in addition to hierarchies with respect to external scales, there can also be internal scales, i.e. scales that depend on loop momenta, that can get small or large with respect to external or other internal scales. An example of such a scale in the present case is the center-of-mass energy of the partonic quark-photon subprocess
\begin{align}
\hat s = \frac{x-\rho}{\rho} Q^2,
\end{align}
which gets small with respect to the hard scale when $x$ is close to $\rho$. Strictly speaking, when we talk about ``threshold'' in the context of this work, we generally refer to the partonic threshold at $\hat s = 0$, as opposed to the hadronic threshold given by the smallest singularity in the center-of-mass energy
\begin{align}
s = (p+q)^2 = \frac{1-\rho}{\rho} Q^2 + m^2 - \frac{t}{4}
\end{align}
at some $s = s_{\rm th}$.
In the case where both photons are off-shell, i.e. $\rho \neq \pm \xi$, we will argue that
\begin{align}
\f F_{\perp}^{\rm ns}(\rho, \xi, t, Q) &= H(-q^2, \mu) H(-q'^2, \mu)   \int_{-1}^1 \frac{dx}{\rho - x - i0} 
\notag
\\
&\quad \times\Big [ \f G\Big ( \frac{\rho - x - i0 }{\rho} Q^2, \mu \Big ) + \hat O(\rho - x) \Big ]  F_q(x,\xi,t,\mu),
\label{eq: x to rho factorization}
\end{align}
where $\hat O$ means big-$O$ up to arbitrary powers of logarithms\footnote{More precisely, $f(x) = \hat O(g(x))$ as $x \rightarrow a$ if there exists $n \in \bb N$ such that $f(x) = O(g(x) \log^n(x-a))$ as $x \rightarrow a$. As usual the phrase ``as $x \rightarrow a$'' is understood with $a$ being clear from the context. It should also be mentioned that the order estimate in eq. \eqref{eq: x to rho factorization} (and also for similar equations that appear in the following) holds only at any fixed order in the perturbative expansion of $\f G$.},
$H$ is the hard matching coefficient of the Sudakov form factor (SFF) and $\f G(-\hat s, \mu)/(-\hat s)$ is the quark propagator in the light-cone gauge. Factorization formulas like eq. \eqref{eq: x to rho factorization} enable one to resum an infinite number of terms by the use of evolution equations for the factors. One can add these all-order in $\alpha_s$ terms to the fixed order unexpanded result to obtain a more ``precise'' result.

Clearly, the expansion in powers of $\rho - x$ in eq. \eqref{eq: x to rho factorization} is not useful for integration regions where $x$ is not close to $\rho$. Moreover, if we can deform the $x$ integration contour away from $\rho$ into the complex plane, it is not a good approximation anywhere on the integration path. Moreover, the logarithms $\log(\rho - x)$ that can be resummed in $\f G$ are not large on the integration contour. Upon closer inspection, it is not hard to see that the partonic threshold region $x \sim \rho$ gives a dominant contribution to the imaginary part of $\f F_{\perp}^{\rm ns}$ in the hadronic threshold region $\rho \sim 1$, as will be explained in section \ref{sec: resum}. This is familiar from the threshold $x_B \rightarrow 1$ resummation in DIS, where the cross section is determined by ${\rm Im} \, \f F_{\perp}^{\rm ns}(x_B , 0,0, Q)$. 

Since the parton densities vanish at $x  =1$, the $x_B \rightarrow 1$ region in DIS is of phenomenologically limited importance and this also the case for off-forward Compton scattering. The resummation is nevertheless conceptually interesting, in particular regarding its simplicity, and it can inform resummations for similar processes where the logarithms are large. At the very least, one can easily predict higher order terms in the fixed order CFs that can be used as checks for future calculations.

Note that eq. \eqref{eq: x to rho factorization} also enables one to resum not only $\log \frac{-\hat s}{Q^2}$ and $\log \frac{-\hat u}{Q^2}$ but also $\log \frac{-q^2}{Q^2} = \log \frac{\xi + \rho}{\rho}$ and $\log \frac{-q'^2}{Q^2} = \log \frac{\xi - \rho}{\rho}$, corresponding to the regions $|q^2| \gg |q'^2| \gg |\hat s|$ or $|q'^2| \gg |q^2 |\gg |\hat s|$, respectively. 

We will (re)derive the factorization formula in eq. \eqref{eq: x to rho factorization} in section \ref{sec: DDVCS fact}. The corresponding result for the imaginary part in the forward case $q = q'$ is well-known \cite{Sterman:1986aj, Catani:1989ne, Korchemsky:1992xv, Becher:2006mr, Chen:2006vd}. The arguments can be for the most part exactly copied in the case that both photons are far off-shell, with the additional simplification that one does not have to take the imaginary part. In this work, we pursue a different argument analogous to that in \cite{Schoenleber:2022myb}.

That is, we view eq. \eqref{eq: x to rho factorization} as a factorization of the CF $C$ itself
\begin{align}
C(x/\rho, \xi/\rho, \mu, Q) = \frac{Q^2}{-\hat s - i0} H(-q^2,\mu) H(-q'^2, \mu)  \f G(-\hat s, \mu) + \hat O((|\hat s|/Q^2)^0).
\label{eq: x to rho factorization for CF}
\end{align}
The expansion for $x \rightarrow - \rho$ is related to eq. \eqref{eq: x to rho factorization for CF} by crossing symmetry $\hat s \leftrightarrow \hat u = -\frac{x + \rho}{\rho} Q^2$. That being said, for the rest of this work we only deal with the $\hat s$-channel, i.e. $x \rightarrow \rho$, the $\hat u$-channel, i.e. $x \rightarrow - \rho$, being completely analogous and obtainable from crossing symmetry.

We remark that in the case where one of the photons, say the final state photon, is on-shell, i.e. DVCS, we get a different factorization near the threshold, namely
\begin{align}
C_{\rm DVCS}(x/\xi, \mu, Q) &= \frac{- q^2}{2(-\hat s - i0)} H(-q^2,\mu)  \f I(-\hat s - i0, \mu) + \hat O((|\hat s|/Q^2)^0),
\label{eq: DVCS thresh fact}
\end{align} 
where $\f I$ is the radiative jet function, which is known to two loop accuracy \cite{Liu:2020ydl}. Eq. \eqref{eq: DVCS thresh fact} was derived in \cite{Schoenleber:2022myb} and then again in \cite{Schoenleber:2024ihr} using soft-collinear effective theory (SCET). Note that it is not possible to recover eq. \eqref{eq: DVCS thresh fact} from eq. \eqref{eq: x to rho factorization for CF}, since the expansions in $\hat s$ and $q'^2$ do not commute. Indeed, while $C_{\rm DVCS} = \lim_{q'^2 \rightarrow 0} C$ is well-defined, this is clearly not the case for the leading power term in $|\hat s|/Q^2$, since $H(-q'^2, \mu^2)$ diverges logarithmically as $q'^2 \rightarrow 0$.

Note that the amplitudes in eq. \eqref{eq: main factorization DDVCS} contribute for transversely polarized photons. If both photons are off-shell there is a leading power contribution to the Compton amplitude where both photons are longitudinally polarized. From our discussion in section \ref{sec: DDVCS fact} it will become clear that this contribution to the hard scattering is subleading in powers of $|\hat s|/Q^2$. This can be readily verified from the one-loop results for the CFs \cite{Mankiewicz:1997bk, Belitsky:2005qn}. Moreover, the results can be trivially extended to the full quark sector, including the pure-singlet contributions that mix with gluons, since these contributions will also turn out to be subleading in powers of $|\hat s|/Q^2$.

This work is structured as follows. In section \ref{sec: prelims} we introduce necessary definitions and conventions. In section \ref{sec: DDVCS fact} we discuss the threshold region of DDVCS and derive the factorization formula \eqref{eq: x to rho factorization for CF}. 
In section \ref{sec: resum} we discuss the resummation of threshold logarithms in DDVCS and in section \ref{sec: DIS} we discuss the special case of DIS, which amounts to taking the imaginary part of eq. \eqref{eq: x to rho factorization} for forward kinematics, recovering the well-known result. Finally, in section \ref{sec: 2loop DDVCS at LO} we present the two-loop quark CF of DDVCS at leading power in $|\hat s|/Q^2$, which has been found to agree with the recent result \cite{Braun:2024srt} from explicit calculation.

\section{Preliminaries}
\label{sec: prelims}
We define
\begin{align}
P &= \frac{p+p'}{2}, \qquad t = (p'-p)^2, \qquad m^2 = p^2 = p'^2
\\
Q^2 &= - \frac{1}{4} (q+q')^2, \qquad \rho = - \frac{(q+q')^2}{4P \cdot (q+q')}, \qquad \xi = - \frac{\Delta \cdot (q+q')}{2P \cdot (q+q')}.
\label{eq: kin defs}
\end{align}
We introduce two light-cone vectors
\begin{align}
n^{\mu} = \frac{1}{\sqrt{2}} (1,0,0,-1)^{\mu},  \qquad \bar n^{\mu} = \frac{1}{\sqrt{2}} (1,0,0,1)^{\mu}.
\end{align}
For a generic vector $V$, we define $V^+ = n \cdot V,~ V^- = \bar n \cdot V$ and $V_{\perp}^\mu = V^\mu - V^+ \bar n^\mu - V^- n^\mu$. We will commonly denote a vector $V$ in terms of its light-cone components by $V = (V^+, V^-, V_{\perp})$.

In the center-of-mass frame, the momenta of the hard quark-photon subprocess at leading power are given by
\begin{align} \notag
\hat p^{\mu} &= (x+\xi) P^+ \bar n^{\mu},
\\ \notag
\hat p'^{\mu} &= (x- \xi) P^+ \bar n^{\mu},
\\ \label{eq: hatted momenta}
\hat q^{\mu} &= - (\rho + \xi) P^+ \bar n^{\mu} + \frac{Q^2}{2\rho P^+} n^{\mu},
\\ \notag
\hat q'^{\mu} &= - (\rho - \xi) P^+ \bar n^{\mu} + \frac{Q^2}{2\rho P^+} n^{\mu}.
\end{align}
As usual, the parton momentum fraction $x$ parametrizes the plus component of the loop momentum that connects the hard and collinear sectors (after using Ward identities to decouple the scalar-polarized gluons from the hard scattering).

It will be convenient to introduce two power counting parameters. Firstly, we define
\begin{align}
\lambda \sim \frac{m}{Q} \sim  \frac{\Lambda_{\rm QCD}}{Q} \sim \frac{\sqrt{-t}}{Q} 
\end{align}
which parametrizes the asymptotic limit $Q \rightarrow \infty$. 
Secondly, we define
\begin{align}
\eta \sim \frac{\sqrt{|\hat s|}}{Q}  = \sqrt{ \frac{|\rho - x|}{\rho} }
\end{align}
as parametrizing the asymptotic approach to the partonic thresholds. Note that $x$ can be viewed as a complex number, since it may be useful to deform the integration contour in eq. \eqref{eq: main factorization DDVCS}, so we point out that $|\rho - x| = \sqrt{(\rho - x^*)(\rho - x)}$, where ${}^*$ denotes complex conjugation.

Throughout this work, we focus on the flavor non-singlet contribution, while the flavor singlet (quark and gluon) contributions are left for future work.

From now when we refer to ``soft'', we mean a generic scaling where all light-cone components of a soft momentum are much smaller than $Q$. It will not be necessary to specify the soft scaling further than that in the present context.

For perturbative quantities $X(\alpha_s)$, we denote the coefficients in the expansion according to
\begin{align}
X = \sum_{n = 0}^{\infty} \Big ( \frac{\alpha_s}{4\pi} \Big )^n X^{(n)}.
\end{align}

\section{All-order factorization for the CF of DDVCS}
\label{sec: DDVCS fact}
\begin{figure}
\centering
a)\includegraphics[scale=.4]{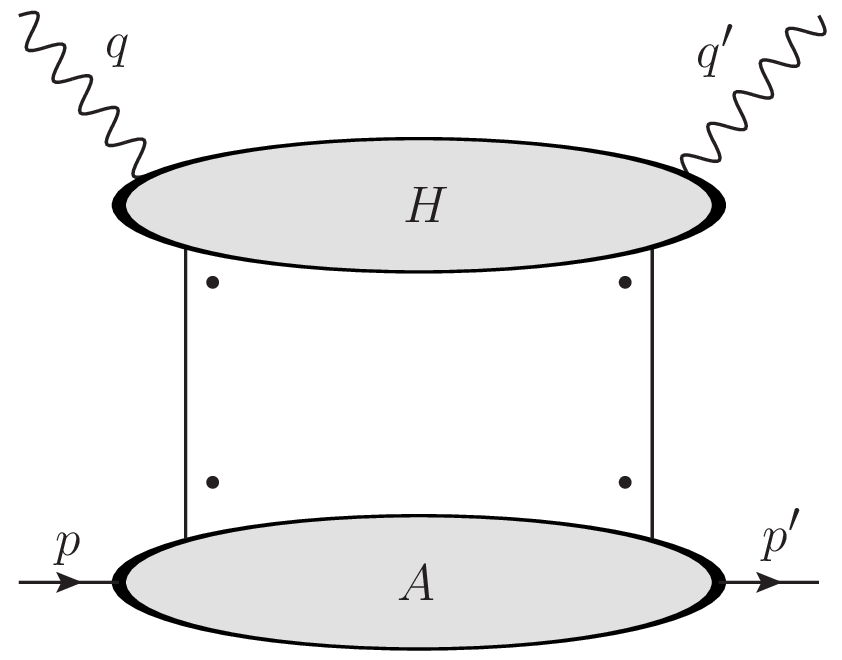}
\hspace{.5cm}
b)\includegraphics[scale=.4]{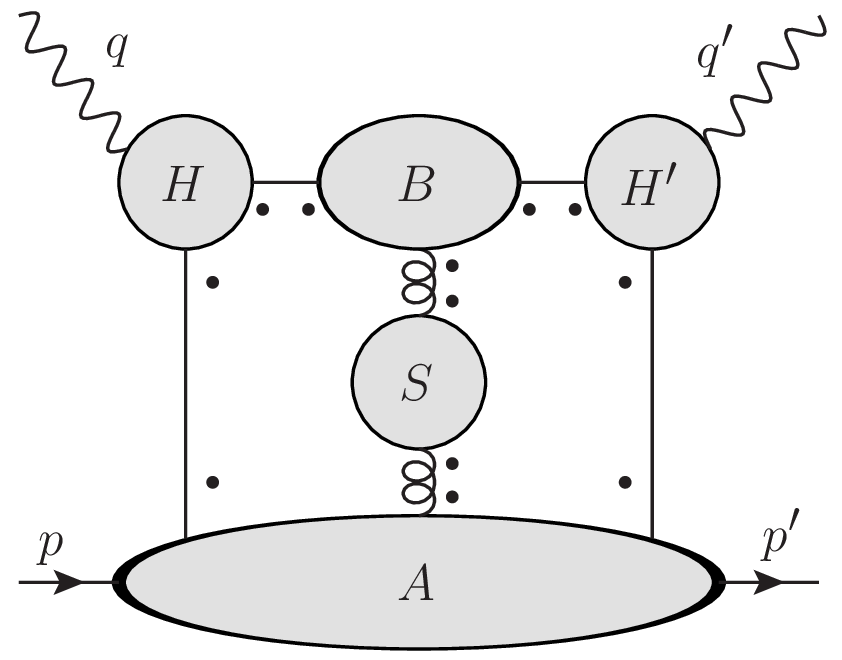}
\caption{a) Leading region for $|x- \rho| \sim 1$ for the virtual Compton process. $A,B$ are $\bar n,n$-collinear subgraphs respectively, $H,H'$ are hard subgraph and $S$ denotes a soft subgraph. b) Leading region as $x \rightarrow \rho$ in DDVCS. Dots beside lines denote an arbitrary number of scalar-polarized gluons.}
\label{fig: DDVCS region}
\end{figure}
\begin{figure}
\centering
a)\includegraphics[scale=.5]{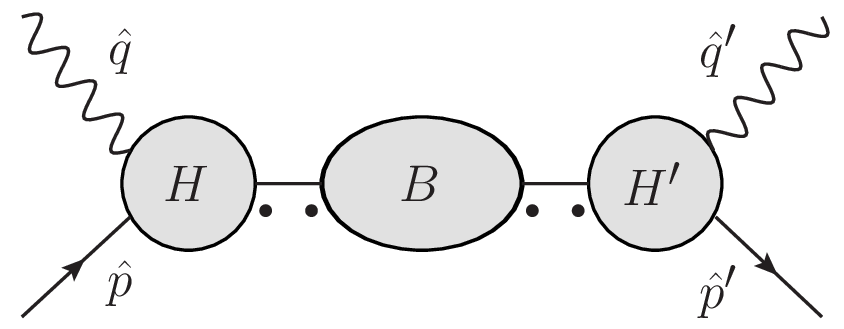}\\ \vspace{.3cm}
b)\includegraphics[scale=.5]{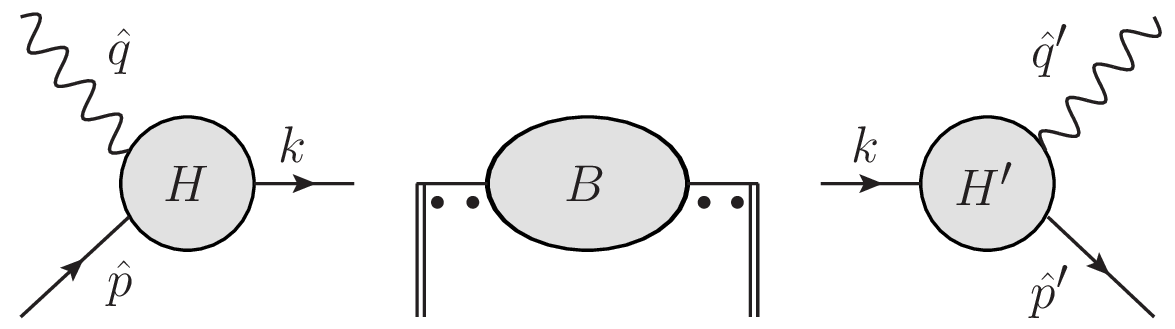}
\caption{a) Leading region as $x \rightarrow \rho$ of $C$. b) Factorised graph after applying the region expansion in $\eta$.} 
\label{fig: DDVCS CF}
\end{figure}
For a generic Feynman graph, we consider regions of loop momentum space, where the invariant mass of the particles produced in the initial parton-photon collision $|\hat s|$ is much smaller than $Q^2$. Consider, as an intuitive example, a single quark of momentum
\begin{align*}
k^{\mu} = \hat p^{\mu} + \hat q^{\mu} = (x-\rho) P^+ \bar n^{\mu} + \frac{Q^2}{2\rho P^+} n^{\mu}
\end{align*} 
that leaves the hard scattering at the incoming photon vertex. $|\hat s| = |k^2| \ll Q^2$, or equivalently $\eta \ll 1$, implies that the quark has a small plus momentum, so that it becomes anti-collinear to the target. In other words, we can view this as the physical process where the parton coming from the target hits the photon, flipping partons direction while it remains approximately on-shell. It then emits the outgoing photon which flips the direction again, to recombine with the target. This corresponds to the $x \rightarrow \rho$ region of the loop momentum space. 

In principle one can have soft interactions between the $n$-collinear (i.e. $k$-collinear) lines and $\bar n$-collinear (i.e. target-collinear) lines. The leading regions as $-\hat s \rightarrow Q^2$ for the complete amplitude are shown in Figure \ref{fig: DDVCS region}b. This is opposed to the situation where the scattered quark has both large plus and minus component, Figure \ref{fig: DDVCS region}a, which leads to eq. \eqref{eq: main factorization DDVCS}. In that case the invariant mass $\hat s$ is large, which means that $x$ can not be close to $\rho$. 

However, the scaling $k \sim (\eta^2, 1, \lambda) Q$ with $\eta \ll 1$ does not correspond to a pinched configuration of loop momenta, that is, there is no actual IR divergence. In other words, there are no propagator denominators that would obstruct a contour deformation making $\eta \sim 1$. For this to be the case we would actually need a collinear external particle attached to the $B$ subgraph. We can therefore deform the contour away from the $\frac{1}{x-\rho}$ pole\footnote{Up to the subteltly regarding the non-analyticity of the GPD, see \cite{Collins:1998be}}. In fact this contour deformation has to be implicitly performed when evaluating the convolution integral in eq. \eqref{eq: main factorization DDVCS}, the direction being given by the Feynman pole prescription, which corresponds to $-\hat s - i0$ or equivalently $\rho - i0$.

One could now proceed, similar to \cite{Becher:2006mr}, by factorising region \ref{fig: DDVCS region}b, which would then lead directly to eq. \eqref{eq: x to rho factorization}. However, as was done in \cite{Schoenleber:2022myb}, it is simpler to use the fact that the CF is an amplitude for massless and on-shell partons that are collinear in the same direction. Then any target-collinear or soft region gives scaleless integrals and one is left with only two regions, hard and $n$-collinear. The factorization with only these two regions is very simple and follows from standard arguments \cite{Collins:2011zzd}. Of course, the CF is not just the bare amplitude, but is subject to double-counting subtractions that subtract the IR divergences due to having massless and on-shell partons. However, given that the ``bare'' (i.e. unsubtracted) amplitude factorizes, the subtractions must cancel (by construction) on both sides of the ``bare'' factorization formula implying the factorization of the subtracted CF \cite{Schoenleber:2022myb}.
 
We consider the leading regions for the CF in Figure \ref{fig: DDVCS CF}a. The CF is defined with the limit $\lambda \rightarrow 0$ applied. This means that the hatted (partonic) momenta in eq. \eqref{eq: hatted momenta} shall be taken as external with respect to the $\hat p + \hat q \rightarrow \hat p' + \hat q'$ subprocess. For a generic loop momentum $l$, the hard region is now characterized by having $l^{\pm} \sim Q$, while the $n$-collinear region has $l^+ \sim \eta^2 Q$ and $l^- \sim Q$. As usual, for hard and collinear loop momenta in the method of regions expansion, the $\perp$ component of loop momenta must be fixed such that $l_{\perp}^2 \sim l^+ l^-$, i.e. $l_{\perp} \sim Q$ for hard and $l_{\perp} \sim \eta Q$ for $n$-collinear. Recall that the hard scale is $q^2, q'^2 \sim Q^2$ and the $n$-collinear scale is $k^2 \sim \eta^2 Q^2 \sim -\hat s$. The $\bar n$-collinear region always leads to scaleless loop integrals \cite{Schoenleber:2022myb}, since the associated scale $\hat p^2, \hat p'^2, \hat p \cdot \hat p' = 0$ is set to zero. The same holds for the soft region, since the soft scale is given by product of the $n$-collinear and $\bar n$-collinear scale (divided by $Q^2$), i.e. zero. I remark that the precise scaling relation between $\eta$ and $\lambda$ is immaterial. What is important is rather the order of expansion, that is: we first expand in $\lambda$ and then expand in $\eta$. This may generally understood by a notation such as $1 \gg \eta \gg \lambda$. This notation might be misleading, however, because $\eta$ is an integration variable that is integrated over a range containing zero. 

We have reduced the problem to two regions, hard and $n$-collinear. The leading region in \ref{fig: DDVCS CF}a follows from the standard power counting formulas \cite{Collins:2011zzd}, while the leading term in the expansion in $\eta$ gives the factorised expression in \ref{fig: DDVCS CF}b, by using standard Ward identity arguments implying that the scalar polarized gluons combine to Wilson lines.
The factor depending on the small scale $-\hat s$, the middle graph in Figure \ref{fig: DDVCS CF}b, after factoring out an overall $\frac{Q^2}{-\hat s}$, reads
\begin{align}
\f G(- \hat s, \mu) = \frac{1}{4N_c} {\rm tr} \Big [ (- i \s k) \int d^dx \, e^{-i k \cdot x} \langle \Omega | {\rm T} \, W_n^{\dagger}(0) \psi(0) \bar \psi(x) W_n(x) | \Omega \rangle \Big ],
\label{eq: J def}
\end{align}
where the trace is over Dirac and color space,
\begin{align}
W_n(x) = {\rm P} \exp \Big [ ig \int_{-\infty}^0 ds\, \bar n \cdot A(x + s \bar n) \Big ]
\end{align}
is a Wilson line in the fundamental representation
and ${\rm T}, \, {\rm P}$ denote time-, path-ordering respectively. For the expression in eq. \eqref{eq: J def} renormalization in dimensional regularization with the $\overline{{\rm MS}}$-scheme is  implied. 

The hard subamplitudes at the photon vertices, stripped of the Dirac structure which is simply $\gamma_{\perp}^{\mu(\nu)}$, is precisely given by the hard matching coefficient of the SFF (after application of double-counting subtractions as described earlier). Since the $n$-collinear subamplitude is proportional to $\gamma^+$, the overall Dirac and Lorentz structure simplifies greatly in the $x \rightarrow \rho$ region. Indeed, since chirality is conserved in the hard scattering, the Dirac trace will always reduce to 
\begin{align}
\frac{1}{4}{\rm tr}\, \gamma^- \gamma_{\perp}^{\mu} \gamma^+ \gamma_{\perp}^{\nu} = - g_{\perp}^{\mu \nu},
\end{align}
where $g_{\perp}^{\mu \nu} = g^{\mu \nu} - n^{\mu} \bar n^{\nu} - n^{\nu} \bar n^{\mu}$,
for the vector contribution and
\begin{align}
\frac{1}{4} {\rm tr}\, \gamma^- \gamma_5 \gamma_{\perp}^{\mu} \gamma^+ \gamma_{\perp}^{\nu} = i\varepsilon^{\mu \nu \rho \sigma} n_{\rho} \bar n_{\sigma}
\end{align}
for the axial-vector contribution. Note that other than that the $\gamma_5$ is immaterial, so we conclude that the leading power contribution in $\eta$ for the vector and axial-vector CF is equal. Note that for longitudinally polarized photons the polarization vector contracted with $\gamma_{\perp}^{\mu}$ gives zero, which means that this contribution is subleading in $\eta$. We conclude that the Compton amplitude with longitudinally polarized photons is $\hat O(\eta^0)$ to all orders in $\alpha_s$.

Furthermore, note that any quark ``box-like'' topologies, meaning graphs with quark lines connecting the initial and final state $\bar n$-collinear lines, lead to a suppression in $\eta$. This is because additional quark connections between collinear and hard subgraphs always imply a suppression by standard power-counting \cite{Collins:2011zzd}. This implies that, as mentioned before in the introduction, that eq. \eqref{eq: main factorization DDVCS} applies also to the full quark coefficient function, including the pure-singlet terms that mix with gluons, since any graph contributing to this sector necessarily contains a quark box and hence additional quark lines connecting $H$ or $H'$ to $B$. We conclude that the pure-singlet quark contribution to the Compton amplitude is also $\hat O(\eta^0)$ to all orders in $\alpha_s$.

Finally, after restoring the Feynman pole prescription by replacing $-\hat s \rightarrow -\hat s - i0$, we obtain the factorization formula in eq. \eqref{eq: x to rho factorization for CF}.

\section{Resummation for the CF of DDVCS}
\label{sec: resum}
In this section we solve the evolution equations for the factors appearing in eq. \eqref{eq: x to rho factorization for CF} and discuss the results.
Instead of $\f G$ it is more common in the literature to consider the function
\begin{align}
\f J(-\hat s, \mu) = \frac{\f G(-\hat s, \mu)}{- \hat s},
\end{align}
which obeys the same evolution equation as $\f G$, namely
\begin{align}
\mu \frac{d}{d\mu} \f J(- \hat s, \mu) &= \left [ -2 \Gamma_{\rm cusp}(\alpha_s(\mu)) \log \frac{-\hat s}{\mu^2} -2 \gamma_{\f J}(\alpha_s(\mu)) \right ] \f J(-\hat s, \mu). 
\label{eq: evo J}
\end{align}
$\f J$ is the quark propagator in axial-gauge and its imaginary part divided by $\pi$ is known as the jet function
\begin{align}
J(\hat s, \mu) = \frac{1}{\pi} {\rm Im}\, \f J(-\hat s - i0, \mu),
\label{eq: jet fun}
\end{align}
which obeys a more complicated evolution equation \cite{Becher:2006qw}.
The evolution equation for the CF of the SFF reads
\begin{align}
\mu \frac{d}{d\mu} H(Q^2, \mu^2) = \left [ \Gamma_{\rm cusp}(\alpha_s(\mu)) \log \frac{Q^2}{\mu^2} + \gamma_H(\alpha_s(\mu)) \right ] H(Q^2, \mu^2). 
\label{eq: evo H}
\end{align} 
Let us now evolve the factors in eq. \eqref{eq: x to rho factorization for CF} to this common scale, giving
\begin{align}
\f C(q^2,q'^2,-\hat s, \mu) &= \exp \left [ -4 S(\mu_i, \mu) +2 a_{\gamma_{\f J} }(\mu_i,\mu) + 2S(\mu_h, \mu) - a_{ \gamma_H}(\mu_h ,\mu) + 2S(\mu_h', \mu) - a_{
\gamma_H}(\mu_h' ,\mu) \right ]  \notag
\\
&\quad \times \left ( \frac{-q^2}{\mu_h^2} \right )^{-\, a_{\Gamma_{\rm cusp}}(\mu_h,\mu) }\left ( \frac{-q'^2}{\mu_h'^2} \right )^{-\, a_{\Gamma_{\rm cusp}}(\mu_h',\mu) } \left ( \frac{-\hat s}{\mu_i^2} \right )^{2 \, a_{\Gamma_{\rm cusp}}(\mu_i,\mu) } \notag
\\
&\quad \times H({-q^2}, \mu_h) H({-q^2}, \mu_h') \f J({-\hat s}, \mu_i)
\label{eq: C resumed}
\end{align}
where
\begin{align}
S(\nu,\mu) &= - \int_{\alpha_s(\nu)}^{\alpha_s(\mu)} d\alpha \, \frac{\Gamma_{\rm cusp}(\alpha)}{\beta(\alpha)} \int^{\alpha}_{\alpha_s(\nu)} \frac{d\alpha'}{\beta(\alpha')},
\\
a_{\gamma}(\nu,\mu) &= - \int_{\alpha_s(\nu)}^{\alpha_s(\mu)} d\alpha\, \frac{\gamma(\alpha)}{\beta(\alpha)}
\end{align}
and $\beta = \frac{d\alpha_s}{d\log \mu}$ is the QCD beta function. 

\noindent For instance, setting $\mu = Q$, and expanding the exponent in $\alpha_s(Q)$, we can observe the exponentiation of the leading double logarithms
\begin{align}
\f C(q^2,q'^2,-\hat s, Q) = \exp \Bigg [  \frac{\alpha_s(Q)}{4\pi} \Big ( 2 \log^2 \frac{-\hat s}{Q^2} - \log^2 \frac{-q^2}{Q^2} - \log^2 \frac{-q'^2}{Q^2} \Big ) + ... \Bigg ].
\end{align}
All the ingredients for the next-to-next-to-leading logarithmic accuracy are known, but, since this resummation is mostly of conceptual interest, we will not pursue a detailed study of the corrections in this work.

In eq. \eqref{eq: C resumed} the choices for the scale $\mu_h, \mu_h'$ and $\mu_i$ should be such that the logarithms are minimized. This is somewhat tricky for $\f J$. The choice $\mu_i^2 = - \hat s$ clearly minimizes the logarithms, but it necessitates integrating over the running coupling $\alpha_s(\sqrt{-\hat s })$. This approach was used in \cite{Schoenleber:2022myb} for DVCS, by integrating around the corresponding essential singularity in the complex $x$ plane. Ambiguities in handling the singularities signify an IR sensitivity of the CF and this becomes particularly pronounced for the behaviour near $\rho \sim 1$. Indeed, while ${\rm Im}\,\f F_{\perp}^{\rm ns}$ vanishes as $\rho \rightarrow 1$ at any fixed order, the Landau singularity produces a finite imaginary part of the amplitude at $\rho = 1$. These ambiguities are of the same parametric order as power-suppressed (in $\lambda$) corrections \cite{Contopanagos:1993yq, Beneke:1995pq}.

Alternatively, the Landau pole can be avoided by choosing $\mu_i$ to be fixed, but this does only make sense if that choice also minimizes the logarithms. In fact this is certainly possible for the imaginary part of $\f F_{\perp}^{\rm ns}$. Indeed, notice that the imaginary part of $\f J$ has support for $|x| > |\rho|$, so we get integrals of the form
\begin{align}
\int_{\rho}^1 dx \, (x-\rho)^n \log^j(x-\rho) \sim (1-\rho)^{n+1} \log^j(1-\rho) \qquad \text{ as }\qquad \rho \rightarrow 1.
\end{align}
This means that the expansion in $x-\rho$ can be identified with the expansion in $1-\rho$, where logarithms $\log(x-\rho)$ correspond to $\log(1-\rho)$. Therefore, for the imaginary part, the natural choice is $\mu_i^2 \sim (1-\rho) Q^2$. 
One may consider using this fixed scale also for the real part.
However, it is easy to see that the partonic threshold expansion in $x-\rho$ can \textit{not} be identified with the expansion of ${\rm Re}\, \f F_{\perp}^{\rm ns}$ in $1-\rho$, since then there are also logs present corresponding to the other endpoint of the $x$ integral which become enhanced for the scale choice $\mu_i^2 \sim (1-\rho) Q^2$. The same issue appears in the threshold resummation for lattice calculations of distribution amplitudes \cite{Baker:2024zcd, Cloet:2024vbv}. 


The real part is determined in terms of the imaginary part by dispersion relations \cite{Diehl:2007jb}, so we could think of calculating first the ``resummation improved'' imaginary part using the fixed scale choice and then recovering the real part from the dispersion relation. But the dispersion relation involves an integral of ${\rm Im}\,\f F_{\perp}^{\rm ns}$ over the range $1 > \rho > 0$ and we can not evaluate ${\rm Im}\,\f F_{\perp}^{\rm ns}$ for arbitrarily small $1-\rho$ because we encounter, once again, the Landau pole. Note that this happens precisely when $\mu_i \sim \Lambda_{\rm QCD}$, where the factorization in eq. \eqref{eq: main factorization DDVCS} is technically invalid. However, because the GPD vanishes as $x \rightarrow 1$, ${\rm Im}\,\f F_{\perp}^{\rm ns}$ also vanishes, so that the endpoint region $\rho \sim 1$ of the dispersion integral should be suppressed, so the region $1-\rho \ll \Lambda_{\rm QCD}^2/Q^2$ of the dispersion integral can potentially be ignored.

A completely analogous considerations apply to the threshold resummation in the DVCS case \cite{Schoenleber:2022myb}, where a simple numerical analysis was performed with the conclusion that corrections due to the resummation are small.
A detailed analysis of these issues and their connection to power corrections would be interesting, but is beyond the scope of this paper and might be pursued in future work.

\section{DIS} \label{sec: DIS}
The cross section for DIS is obtained from the DDVCS amplitude by taking the imaginary part of the forward kinematics $q = q'$. In this case we can identify $\rho = x_B = \frac{Q^2}{2p \cdot q}$. For the factorized expression in eq. \eqref{eq: x to rho factorization} this gives 
\begin{align}
\frac{1}{\pi} {\rm Im} \, \f F_{\perp}^{\rm ns, \,  forward} &= ( H(Q^2, \mu) )^2 Q^2  \int_{x_B}^1 \frac{dx}{x_B}\, J( \hat s, \mu ) f_q^{\rm val}(x, \mu) + \hat O((1-x_B) f_q^{\rm val}(x_B,\mu)) ,
\label{eq: DIS refactorization}
\end{align}
where $f_q^{\rm val}$ is the valence quark PDF and $J$ was defined in eq. \eqref{eq: jet fun}. As explained in section \ref{sec: resum}, for the imaginary part of the amplitude, the partonic threshold expansion in $x-x_B$ corresponds to the actual threshold expansion in $1-x_B$ and we have used this fact in writing eq. \eqref{eq: DIS refactorization}.

The resummation for DIS becomes somewhat more complicated than in the DDVCS case\footnote{Note that taking the imaginary analytically can be avoided, by alternatively performing the countour integral for $\f F_{\perp}^{\rm ns, \,  forward}$ numerically and then extracting the imaginary part.}, since $J$ is distribution-valued, for instance $J^{(0)}(-\hat s) = \delta(-\hat s)$. To simplify the situation one can consider moments
\begin{align}
J_N(Q^2, \mu) = \int_0^{Q^2} dz \, \left ( 1- \frac{z}{Q^2} \right )^{N-1} J(z,\mu),
\end{align} 
but it is even more elegant to consider, following \cite{Becher:2006mr}, the Laplace transform
\begin{align}
\w j\left ( \log \frac{\nu^2}{\mu^2}, \mu \right ) = \int_0^{\infty} dz \, \exp \left ( - \frac{z}{e^{\gamma_E} \nu^2} \right ) J(z,\mu),
\end{align}
which is known as the associated jet function. The evolution equation for $\w j$ then has the same form as the one for $\f J$, namely
\begin{align}
\mu \frac{d}{d\mu} \w j(L,\mu) = \left [ - 2 \Gamma_{\rm cusp}(\alpha_s(\mu)) L - 2 \gamma_J(\alpha_s(\mu)) \right ] \w j(L, \mu). 
\end{align} 
Moreover, the function $\w j$ can be readily related to $\f J$. To see this, consider their expansions
\begin{align}
\f J(-\hat s, \mu) &=  \sum_{n=0}^{\infty} \Big ( \frac{\alpha_s}{4\pi} \Big )^n \sum_{m=0}^{2n} c_{nm} \frac{1}{-\hat s} \log^m \frac{-\hat s}{\mu^2}
\\
\w j(L, \mu) &= \sum_{n=0}^{\infty} \Big ( \frac{\alpha_s}{4\pi} \Big )^n \sum_{m=0}^{2n} \w c_{nm} L^m.
\end{align}
A straightforward calculation shows that the coefficients of these expansions are related by
\begin{align}
\w c_{nm} &= \sum_{k = 0}^{2n} c_{nk} \lim_{\alpha \rightarrow 0^+} \frac{d^k}{d\alpha^k} \alpha^m \frac{\sin(\pi \alpha)}{\pi}  \frac{e^{\alpha  \gamma_E}  \Gamma(\alpha)}{\Gamma(m+1)}. 
\label{eq: cnm relation}
\end{align}
In the following section, we give the result for $\f J$ in eq. \eqref{eq: J result}, which was obtained from eq. \eqref{eq: cnm relation} and the two-loop result for $\w j$ in \cite{Becher:2006mr}.

\section{Two-loop DDVCS at leading order in $\eta$}
\label{sec: 2loop DDVCS at LO}
Consider the leading coefficient $\f C$ in the $\eta$ expansion
\begin{align}
C = \frac{Q^2}{-\hat s} \, \f C  + \hat O(\eta^0),
\end{align}
which, at one-loop accuracy, is given by
\begin{align}
\f C &= 1 +  \frac{\alpha_s C_F}{4\pi} \Bigg \{ \log \frac{Q^2}{\mu^2} \left ( 4 \log \frac{-\hat s}{Q^2} - 2 \log \frac{-q^2}{Q^2} - 2\log \frac{-q'^2}{Q^2} + 3 \right ) \notag
\\
&\quad + 2 \log^2 \frac{-\hat s}{Q^2} - 3 \log \frac{-\hat s}{Q^2} - \log^2 \frac{-q^2}{Q^2}  + 3 \log \frac{-q^2}{Q^2} - \log^2 \frac{-q'^2}{Q^2} + 3\log \frac{-q'^2}{Q^2}  - 9 \Bigg \} + O(\alpha_s^2).
\end{align}
Written in this form, it is obvious that this function factorizes to this order and the evolution eqs. \eqref{eq: evo J} and \eqref{eq: evo H} can be easily verified. Moreover, $H$ and $J$ are known to two-loop orders \cite{Becher:2006mr} and the function $\f J$ and can be obtained directly from $J$, see section \ref{sec: DIS}. This allows us to determine $\f C^{(2)}$ without explicit calculation.

\noindent First, we present the two-loop expression for $\f J$. 
\begin{align}
\f J(-\hat s, \mu) &= 1 + \Big ( \frac{\alpha_s}{4\pi} \Big ) \f J^{(1)}  + \Big ( \frac{\alpha_s}{4\pi} \Big )^2 \Big [ C_F^2 \f J_F^{(2)} + C_F C_A \f J_A^{(2)} + \beta_0 C_F \f J_{\beta_0}^{(2)} \Big ] + O(\alpha_s^2),\notag
\\[.1cm] 
\f J^{(1)} &= C_F \Big ( 2 L^2-3 L-\frac{\pi ^2}{3}+7 \Big ),  \notag
\\
\f J_F^{(2)} &= 2 L^4-6 L^3+\left(\frac{37}{2}+\frac{2 \pi ^2}{3}\right)
   L^2+\left(-8 \zeta _3+\pi ^2-\frac{45}{2}\right) L-18 \zeta
   _3+\frac{34 \pi ^4}{45}-5 \pi ^2+\frac{205}{8}, \notag
\\
\f J_A^{(2)} &= \left(\frac{8}{3}-\frac{2 \pi ^2}{3}\right) L^2+\left(40
   \zeta _3-\frac{73}{9}\right) L-18 \zeta _3-\frac{19 \pi
   ^4}{60}+\frac{\pi ^2}{9}+\frac{1417}{108}, \notag
\\
\f J_{\beta_0}^{(2)} &= -\frac{2 L^3}{3}+\frac{29 L^2}{6}-\frac{247 L}{18}-\frac{4
   \zeta _3}{3}-\frac{5 \pi ^2}{18}+\frac{4057}{216}, \label{eq: J result}
\end{align}
where $L = \log \frac{-\hat s}{\mu^2}$.

\noindent Using the result for $H$ in \cite{Becher:2006mr} and eq. \eqref{eq: x to rho factorization for CF}, we obtain
\begin{align}
\label{eq: C2 pred}
\f C^{(2)}(\mu = Q) &= C_F^2 \f C_F^{(2)} + C_F C_A \f C_A^{(2)} + \beta_0 C_F \f C_{\beta_0}^{(2)},  \notag
\\ \notag
\f C_F^{(2)} &= 2 L^4-6 L^3+\left(-2 L_1^2+6 L_1-2 L_2^2+6 L_2+\frac{4 \pi
   ^2}{3}-\frac{27}{2}\right) L^2
\\ \notag
&\quad +\left(3 L_1^2-9 L_1+3 L_2^2-9
   L_2-8 \zeta _3+\frac{51}{2}\right)
   L+\frac{L_1^4}{2}+\frac{L_2^4}{2}-3 L_1^3-3 L_2^3
\\ \notag
&\quad +\frac{27
   L_1^2}{2}+L_1^2 L_2^2-3 L_1 L_2^2+\frac{27 L_2^2}{2}-2 \pi
   ^2 L_1
\\ \notag
&\quad -\frac{51 L_1}{2}-3 L_1^2 L_2+9 L_1 L_2-2 \pi ^2
   L_2-\frac{51 L_2}{2}
\\ \notag
&\quad +24 L_1 \zeta _3+24 L_2 \zeta _3-78
   \zeta _3+\frac{19 \pi ^4}{90}+7 \pi ^2+\frac{331}{8},
\\ \notag
\f C_A^{(2)} &= \left(\frac{8}{3}-\frac{2 \pi ^2}{3}\right) L^2+\left(40 \zeta
   _3-\frac{73}{9}\right) L+\left(-\frac{4}{3}+\frac{\pi
   ^2}{3}\right) L_1^2+\left(-\frac{4}{3}+\frac{\pi
   ^2}{3}\right) L_2^2
\\ \notag
&\quad +L_1 \left(\frac{41}{9}-26 \zeta
   _3\right)+L_2 \left(\frac{41}{9}-26 \zeta _3\right)+54 \zeta
   _3+\frac{31 \pi ^4}{180}-\frac{13 \pi ^2}{9}-\frac{73}{12},
\\ \notag
\f C_{\beta_0}^{(2)} &= -\frac{2 L^3}{3}+\frac{29 L^2}{6}-\frac{247
   L}{18}+\frac{L_1^3}{3}+\frac{L_2^3}{3}-\frac{19
   L_1^2}{6}-\frac{19 L_2^2}{6}+\left(\frac{209}{18}+\frac{\pi
   ^2}{3}\right) L_1
\\
&\quad +\left(\frac{209}{18}+\frac{\pi
   ^2}{3}\right) L_2 -2 \zeta _3-\frac{14 \pi
   ^2}{9}-\frac{457}{24},
\end{align}
where
\begin{align}
L = \log \frac{-\hat s}{Q^2} = \log \frac{\rho - x}{\rho}, \qquad L_1 = \frac{-q^2}{Q^2} = \log \frac{\rho + \xi}{\rho}, \qquad L_2 = \frac{-q'^2}{Q^2} = \log \frac{\rho - \xi}{\rho}.
\end{align}
This result can has been cross-checked with the full two-loop calculation of $C$ \cite{Braun:2024srt}.

\section{Conclusions and Outlook}

In this work we have explored the (re)factorization of the CF of DDVCS, which corresponds to a separation of photon virtuality $Q^2$ and the partonic invariant mass $\hat s$. The multiplicative factorization in terms of single scale quantities is remarkably simple in momentum space. 
It leads naturally to the well-known threshold factorization in DIS in the $x_B \rightarrow 1$ limit, by taking the imaginary part of the forward kinematics.

The phenomenological impact of these results is probably negligible, but it is conceptually interesting as an application of factorization to purely perturbative quantities, which can be used to resum logarithms to all orders. As byproducts, one can obtain fixed order expansions of the perturbative kernels without explicit calculations, such as the two-loop quark CF of DDVCS at leading power in $\eta$, given in eq. \eqref{eq: C2 pred}. Therefore, it can also be used a tool to check multi-loop computations.

Interesting directions of research include the application of the same rationale to other processes, where the corrections due to threshold resummation might be significant. These include meson production at threshold (e.g. $J/\psi, \Upsilon$ or $\phi$) and the threshold resummation for quasi- and pseudo- GPDs (for the corresponding analysis in the forward case, see \cite{Ji:2023pba, Ji:2024hit}, and for the distribution amplitude case, see \cite{Baker:2024zcd, Cloet:2024vbv}). 
However, in order to apply the threshold resummation correctly a detailed investigation of the issues regarding choice of the intermediate scale $\mu_i$, discussed at the end of section \ref{sec: resum}, is necessary.

\section*{Acknowledgements}
I thank Vladimir Braun, Xiangdong Ji and Swagato Mukherjee for discussions.
This work was supported by the U.S. Department of Energy through Contract No. DE-SC0012704 and by Laboratory Directed Research and Development (LDRD) funds from Brookhaven Science Associates




\bibliographystyle{apsrev}
\bibliography{bibl}%

\end{document}